\documentclass{pasj00}
\begin{document}
\SetRunningHead{Tamura et al.}{IR Polarimetry of NGC 2071}
\Received{2006/9/11}%{yyyy/mm/dd}
\Accepted{2006/12/26}%{yyyy/mm/dd}

\title{Near-Infrared Imaging Polarimetry of the NGC 2071 Star Forming Region with SIRPOL}

%%% begin:list of authors
\author{Motohide \textsc{TAMURA},\altaffilmark{1, 2}  
        Ryo \textsc{KANDORI},\altaffilmark{1}
        Jun \textsc{HASHIMOTO},\altaffilmark{1, 3}
        Nobuhiko \textsc{KUSAKABE}\altaffilmark{2}}
\author{Yasushi \textsc{NAKAJIMA},\altaffilmark{1} 
        Shuji \textsc{SATO},\altaffilmark{4} 
		Chie \textsc{NAGASHIMA},\altaffilmark{3} 
        Mikio \textsc{KURITA}\altaffilmark{4} }
\author{Tetsuya \textsc{NAGATA},\altaffilmark{5} 
        Takahiro \textsc{NAGAYAMA},\altaffilmark{5} 
        Jim \textsc{HOUGH}\altaffilmark{6}}
\author{Tomoaki \textsc{MATSUMOTO}\altaffilmark{7}, Antonio \textsc{CHRYSOSTOMOU}\altaffilmark{6} }
%\author{Accepted for publication in PASJ.}

\altaffiltext{1}{National Astronomical Observatory, 2-21-1 Osawa, Mitaka, Tokyo 181-8588; 
		   hide@subaru.naoj.org}
\altaffiltext{2}{Graduate University of Advanced Science, 2-21-1 Osawa, Mitaka, Tokyo 181-8588}
\altaffiltext{3}{Department of Physics, Tokyo University of Science, 1-3, Kagurazaka, Shinjuku-ku, Tokyo 162-8601}
\altaffiltext{4}{Department of Astrophysics, Nagoya University, Chikusa-ku, Nagoya 464-8602}
\altaffiltext{5}{Department of Astronomy, Kyoto University, Sakyo-ku, Kyoto 606-8502}
\altaffiltext{6}{Centre for Astrophysics Research, University of Hertfordshire, Hatfield, Herts AL10 9AB, UK}
\altaffiltext{7}{Department of Humanity and Environment, Hosei University, Fujimi, Chiyoda-ku, Tokyo 102-8160}
%\altaffiltext{8}{TBD, and this manuscript is the 6th version.}

%% `\KeyWords{}' always has to be placed before `\maketitle'.
 %Do NOT move this preamble from here!

\maketitle

\begin{abstract}
We have conducted deep $JHKs$ imaging polarimetry of a $\sim$8$'$ $\times$ 8$'$ area 
of the NGC 2071 star forming region. 
Our polarization data have revealed various infrared reflection 
nebulae (IRNe) associated with the central IR young star cluster NGC2071IR and 
identified their illuminating sources. There are at least 4 IRNe in NGC2071IR and 
several additional IRNe are identified around nearby young stars in the same field-of-view.
Each illuminating source coincides with a known near-IR source except for
IRS3, which is only a part of IRN2 and is illuminated by the radio source 1c. 
%Polarization vector images of NGC 2071IR and aperture polarimetry of its cluster
%sources are presented and used for the identification of illuminating source for
%each IRN; 
%we have suggested that IRS3 is simply a part of
%IRN illuminated by a (near-IR invisible) radio source NGC 2071 1c.
Aperture polarimetry of each cluster source is used
to detect
%for the discussion
%on the presence and its orientation of 
unresolved circumstellar
disk/outflow systems. 
Aperture polarimetry of 
the other point-like sources within the field is
made in this region for the first time.
The magnetic field structures 
%(from 7$'$=0.8 pc down to 10$''$=0.02 pc 
%[or, a few arcsec $\sim$ 0.01pc]) 
(from $\sim$1 pc down to $\sim$0.1 pc)
are derived using both aperture polarimetry of the point-like sources
and imaging 
polarimetry of the shocked H$_2$ emission that is seen as the dominant knotty nebulae 
in the $Ks$ band image; they are both of dichroic origin and
the derived field {directions are}
consistent with each other.
The magnetic field direction projected on the sky 
is also consistent with that inferred from the 850 $\mu$m thermal
continuum emission polarimetry
of the central 0.2 pc region, 
but running roughly perpendicular ($\sim$75$\arcdeg$) to the direction of the large scale outflow. 
We argue that 
the field strength is too weak to align the outflow in the large scale field direction
via magnetic braking.
\end{abstract}

\section{Introduction}
SIRPOL is the polarimetry mode of the $JHKs$ band simultaneous infrared (IR) camera 
SIRIUS (Nagashima et al. 1999; Nagayama et al. 2003) 
mounted on the 1.4 m IRSF survey-dedicated telescope 
in South Africa, which has been available since 2005 December. 
The instrument is 
the first to provide deep and wide-field infrared polarimetric images, 
and can in principle measure polarizations of all the 2MASS-detected sources 
within a field-of-view of 7$\farcm$7 $\times$ 7$\farcm$7 in the $JHKs$ bands simultaneously 
with 1\% polarization accuracy. Such a capability is unique and has vast and 
unique applications in many fields in astronomy, including efficient detection 
of circumstellar material that produces scattering (see Weintraub, Goodman, \& Akeson 2000; 
Tamura \& Fukagawa 2005 for a review), and delineating magnetic field 
structures in molecular clouds (e.g., Tamura et al. 1987) and in galaxies 
(e.g., Jones 2005), both through dichroic polarization. 

Wide-field polarimetry is complimentary to the high resolution polarimetry 
with adaptive optics on current large telescopes (e.g., Perrin et al. 2004), 
and highly-polarized sources detected with the wide-field polarimetry 
could provide the best targets for future high resolution telescopes 
at various wavelengths such as ALMA or 30-m class optical-IR telescopes. 
In this paper, we present imaging polarimetry of the NGC 2071 star forming 
region in Orion.

The NGC 2071 star forming region lies in the northern part of the Orion B 
molecular cloud (Lynds 1630) at a distance of 390 pc (Anthony-Twarog 1982).
A dense molecular and dust cloud ``core'' is observed
(e.g., Takano et al. 1986; Dent et al. 1989).
There is a cluster of IR sources in this region, NGC 2071IR, and at least 
9 near-IR and mid-IR sources have been detected in the core region 
(Persson et al. 1981; Lada et al. 1991; Walther et al. 1993).
A red near-IR source (SSV41) originally found with a single-detector raster scan
of a 31$''$ beam by Strom, Strom, \& Vrba (1976)
seems to correspond to
%either one of these or 
some of these collectively.
Snell \& Bally (1986) detected three radio continuum sources, some of which
(NGC 2071 1a and 1b) are likely to have counterparts to the near-IR sources.
The total infrared luminosity of the core region is estimated to be $\sim$520 {$L$}{$_{\odot}$} 
(Butner et al. 1990), which suggests formation of intermediate-mass stars. 

An energetic bipolar CO outflow is centered near NGC2071IR, lying 
in the NE-SW direction 
(see Moriarty-Schieven, Hughes, \& Snell 1989 and references therein). 
Molecular hydrogen emission is also detected (Eisl{\"o}ffel 2000; 
Aspin, Sandell, \& Walther 1992 
and references therein), which is well correlated with the CO outflow 
morphology. Dense gases associated with either outflows or disks are 
observed in CS, SO, SiO, HCO$^{+}$, and NH$_{3}$ (Kitamura et al. 1992; 
Chernin \& Masson 1993; Girart et al. 1999; Zhou, Evans, \& Mundy 1990). 
Several water masers and an OH maser are detected (e.g., Sandell et al. 1985).
Water maser emission mapping with VLBA (Seth et al. 2002)
suggests that IRS1 and IRS3 are the exciting sources
and both sources have compact ($\sim$10 AU) disks.

The region has been one of the first targets of IR polarimetry. 
Aperture polarimetry of IRS1 and its mapping were conducted by Heckert \& 
Zeilik (1984).
They detected a relatively large ($\sim$7\% in the $K$ band 
at $\theta$=142$\arcdeg$) polarization perpendicular to the outflow 
direction with a large 32$''$ beam. Hodapp (1984) and Sato et al. (1985) 
suggested the presence of IR reflection nebulae (IRNe) extending roughly 
in the direction of the large scale outflow. 
Therefore, the aperture polarization with a large beam
can be explained with the dust scattering in the IRN.
More recently, Walther et al. (1993) 
presented high resolution (0$\farcs$62 pixel$^{-1}$) $K$ band imaging polarimetry of a $\sim$65$''$ $\times$ 40$''$ 
region of NGC2071IR, including IRS1 to IRS8a and b, as well as $JHK$ \& $NBL$ band imaging 
of a $\sim$40$''$ $\times$ 35$''$ region and $K$ band imaging of a 5$'$ $\times$ 9$'$ region. 
However, no wide-field, deep imaging polarimetry or $JH$ band polarimetry has 
been conducted yet. 

Information regarding magnetic fields in the NGC 2071 region was obtained 
from 850 $\micron$ polarimetry (Matthews, Fiege, \& Moriarty-Schieven 2002),
which suggests 
the presence of a weakly aligned magnetic field over $\sim$2 pc scale
whose direction is 
roughly perpendicular to the outflow direction. 
A very large scale (15$\arcdeg$ $\sim$ 100 pc) magnetic field structure 
including the Orion molecular cloud complex was studied with 
optical interstellar polarimetry (Appenzeller 1974), 
but their field star polarimetry data are 
too sparse to discuss the magnetic field structure 
in the NGC 2071 region ($<$ 1 pc scale).

%\noindent IMPORTANT NOTICE\\
%1. ``\verb|\draft|'' creates single column and double spaces format.\\
%2. If you comment out ``\verb|\draft|'', the output will be double column
%   and single space.\\
%3. For cross-references, the use of ``\verb|\label|, \verb|\ref|, \verb|\cite|'' 
%   and the thebibliography environment is strongly recommended. \\
%4. Do NOT use ``\verb|\def|, \verb|\renewcommand|''.\\
%5. Do NOT redifine commands provided by PASJ00.cls.\\

%\newpage

\section{Observations}

Observations of the NGC 2071 region were made on 2006 February 23. 
The 1.25 ($J$ band), 1.63 ($H$ band), and 2.14 ($Ks$ band) $\mu$m imaging polarimetry data 
were obtained simultaneously with the SIRIUS camera
and its polarimeter on the 1.4-m IRSF telescope in South Africa.
The SIRIUS camera can
provide $JHKs$ images simultaneously, with 
a field of view
of 7$\farcm$7 $\times$ 7$\farcm$7 and a pixel scale of 0$\farcs$45 pixel$^{-1}$.
The camera is equipped with three 1024 $\times$ 1024 HgCdTe (HAWAII) arrays,
three broad-band filters, and two beam splitters for the simultaneous imaging.
The polarimeter is composed of an achromatic (0.5 - 2.5 $\mu$m) wave-plate 
rotator unit and a high-extinction-ratio polarizer,
both of which are located upstream of the camera and 
at 
{room temperature}.
See Kandori et al. (2006a) for the details of the polarimeter. 

The total integration time was 900 s per wave-plate angle.
We performed 10 s exposures at 4 wave-plate angles 
(0$\arcdeg$, 45$\arcdeg$, 22$\arcdeg$.5, and 67$\arcdeg$.5) 
at 10 dithered positions (1 set). 
The same observation sets were repeated toward the object (9 times) 
and sky coordinates. 
The sky condition was excellent and 
the seeing size during the observations was 0$\farcs$9 (2 pixels) 
in the $J$ band. 

After image calibrations in the standard manner using IRAF (dark subtraction, 
flat-fielding with twilight-flats, bad-pixel substitution, and sky subtraction), 
the Stokes parameters $(I,\, Q,\, U)$, the degree of polarization $P$, and 
the polarization position angle $\theta$ were calculated as follows:  
\[ Q = I_{0} - I_{45}, \ \  U = I_{22.5} - I_{67.5},\]
\[I = (I_{0} + I_{45} + I_{22.5} + I_{67.5} ) / 2,\] 
%\[ I_{\rm disk} = I_{\rm disk + PSF} - I_{\rm PSF},\] 
\[ P = \sqrt{Q^2 + U^2}/I, {\rm ~and} \ \ \theta = (1/2)\, \arctan (U/Q) \ \ .
\]

The accuracy of polarization position angle ($\delta$$\theta$) was estimated to be better than
3$\arcdeg$ based on the measurements during the commisionning run.
The polarization efficiencies at $JHKs$ are also higher than 96\%; thus
no corrcetion was made on the polarization degree ($P$).

%We carried out software aperture polarimetry of point-like sources 
%on the combined intensity images for each wave plate angle. 
%This is because the center of point sources (i.e., aperture center)
%can not be determined satisfactory on the Q and U images. 
%We discuss the result in H band where nebula
%contamination is less than in J band and scattering efficiency is higher than in Ks band. 
%The aperture radius was 3 pixels.
%The Stokes parameters for each source were then derived as described above.

%Point sources on each angle image has slightly different PSF due to varying seeing size. We then used different
%apertures for each image in order to measure the same fraction of stellar flux fallen in the aperture. The adopted
%aperture sizes were 3.00, 3.24, 3.08, and 3.27 pixels for I0? , I22.5? , I45? , and I67.5? , respectively. The sky annulus
%was set to 10 pixel with 5 pixel width. Since polarization degree P is a positive quantity, derived P values tend
%to be overestimated especially for low S/N sources with small polarization. To correct the bias we calculated the
%debiased P using Pdb = ãP2 ? ƒÂP2, where ƒÂP is the error in P (Wardle & Kronberg 1974).
%\begin{figure}
%  \begin{center}
%    \FigureFile(80mm,80mm){fig1.eps}
%    %%% \FigureFile(width,height){filename}
%  \end{center}
%  \caption{This is the first figure.}\label{fig:sample}
%\end{figure}

\section{Results and Discussion}

%In the theory (\cite{key-1})..........

\subsection{Intensity and polarization properties in the overall NGC 2071 region}

%In the theory (\cite{key-1})..........
Figures 1 (a) and (b) show the $JHKs$ color-composite $I$ image and 
the $JHKs$ color-composite $PI$ image in two different scales, respectively. 
Figure 1 (c) summarizes the identification of notable features
discussed in this section.

To the southern part of the $I$ image, there is diffuse near-IR emission 
associated with the optical reflection nebula NGC 2071.
The illuminating source of the nebula, HDE 290861, is just outside of the
image. 

The large, asymmetric bipolar nebula is centered in the field
($\sim$4$'$ north of the optical nebula), 
in which the most striking {feature} is
reddish knotty 
%H$_{2}$ 
{emission in the $Ks$ band.
From the similarlity between our $Ks$ band image and previous H$_{2}$ emission
images (Eisl{\"o}ffel 2000; Aspin et al. 1992; Garden, Russell, \& Burton 1990),
we consider that the emission in our $Ks$ band image comes mostly 
from the H$_{2}$ emission.}
The $Ks$ band emission continuously extends from near IRS1 to the north-east 
direction. There is a gap of the $Ks$ band emission in the south-west, 
and knotty H$_{2}$ emission again appears $\sim$3$'$ to the south-west 
(NGC 2071SW; Walter et al. 1993). 

In contrast, $JH$ emission is more smoothly distributed and forms 
a U-shaped nebula centered near IRS1 extending to the north-east. 
It traces some arc structure in the southern part of the NGC 2071SW region. 
There is also a gap of the $JH$ emission to the south-west. 

Although both of the $Ks$ band emission and the $JH$ band emission 
show the asymmetric bipolar structure, their detailed distributions 
are different, as seen from the color difference in the $I$ image. 
If we overlap these two emission components, the bipolar structure 
is more clearly recognized (as in figure 1a), and their distribution 
is very well correlated with that of the molecular bipolar outflows 
such as the HCO$^{+}$ outflow (Girart et al. 1999). 
Two near-IR emission peaks in the north-east coincide with 
the blueshifted HCO$^{+}$ emission peaks (B1 and B2) and 
those in the south-west with the redshifted peaks R1 and R2. 
Therefore, it is natural to consider that the molecular 
outflows and the near-IR emission or reflection nebulae 
in the core region are related to the large scale outflow activity.
The high spatial resolution Fabry-Perot H$_{2}$ observations 
of Aspin et al. (1992)
indicate that IRS1 is the likely source for the
large scale outflow (see also \S3.2).

The appearance of these nebulae recognized in the $I$ images
is drastically different 
in the $PI$ image. 
First, the diffuse nebula associated with the optical 
reflection nebula NGC2071 disappears, suggesting that 
their polarization is low at near-IR wavelengths. 
This is due to the contribution of near-IR continuum
from small dust grains (Sellgren 1984).
Second, the $Ks$ band nebulae associated with the bipolar 
outflow cannot be seen, suggesting that their polarization 
is also low. 
This is apparently due to the contribution of
H$_{2}$ emission (see \S3.3).
Third and in contrast, several highly-polarized 
nebulae become distinct. 
There are at least 8 such nebulae in the $I$ image.
They are most likely reflection nebulae, thus
referred as IRN(e) in figures 1. 
Some of them (IRN4, 7) are not distinct in the $I$ image.

The illuminating sources for IRN1-4 in the core region are discussed in detail below.
Those for IRN5, IRN6, and IRN9 are
SSV37 (Strom et al. 1976), IRS18 (Walther et al. 1993), and SSV40
(Strom et al. 1976), respectively.
Those for IRN7 and IRN8 are new point-like sources seen in our 
intensity images.
 
Nine IR sources (IRS1-8a, and 8b) in the core region observed by Walther et al. 
(1993) have been confirmed in our three-color composite 
image of the central cluster NGC2071IR 
although some of them are simply a part of near-IR nebula and not
distinct in our images
(see figure 2 for their identification). Their relative positions, and 
aperture photometry and polarimetry are summarized in Table 1.
{Note that the errors in the position angles ($\theta$) include
the uncertainty in the calibrations (see $\S$2)}.
The positional offset in the table is relative to the 2MASS position of IRS1 
(05$^{h}$47$^{m}$4$\fs$77, 00$\arcdeg$21$'$42$\farcs$8; J2000)
with an accuracy of $\sim$0$\farcs$1.
Photometric calibration was made with using the 2MASS-PSC.

There are many other point-like sources in our intensity images 
which have not been
reported before. We did not identify or list each source in this paper, 
but the calibrated FITS images are available upon request.

\subsection{Polarization images of the central NGC2071IR region}

{Figure 2 shows} the $KsHJ$ polarization vector maps of 
the central 61$''$ $\times$ 45$''$ region.  
{Identifications with known IRS are} also indicated.

There are several IRNe associated with the central star cluster NGC2071IR. 
As has been suggested 
by Walther et al. (1993), {prominent} IRNe 
in the $Ks$ band are those around IRS1 and IRS8 with 
a centro-symmetric vector pattern.
{We also note another prominent pattern around IRS4.}
Our $Ks$ band polarization vector map is mostly consistent 
with the $K$ band polarization vector map of Walther et al. (1993).
However, there are both similarities and differences between 
those two vector images in details, which we compare below.
A summary of these discussions is given at the end of this subsection.

%\begin{enumerate}

\subsubsection{IRS1}
The vector pattern around IRS1 is consistent between the two data. 
IRS1 itself is highly polarized (26\% at $Ks$) and the polarization vectors show 
a gradual direction change over this source
(like the character ``S'' from north to south). 
A large part of the IRN associated with NGC2071IR
(whose extent is more than 30$''$) is 
illuminated by IRS1, thus we refer to it as IRN1. 
This corresponds to the white nebula in the $PI$ image (figure 1b). 
The levels of polarization degree of IRN1 range from 25\% to 45\% at $Ks$. 
IRS1 positionally coincides with the radio continuum source NGC 2071 1a (Snell \& Bally 1986) 
and is considered to be self-luminous.

The brightest part of IRN1 is situated just to the east of IRS1, 
whose direction is rather oblique to the large scale outflow direction. 
However, the axis (79$\arcdeg$) of the compact ($\sim$10 AU size)
disk suggested 
from the distribution of H$_{2}$O {masers} around IRS1 (Seth et al. 2002) 
is consistent with the main extension of IRN1 relative to IRS1.
The position angle of our software aperture polarimetry (Table 1)
of IRS1 ($\theta$ = 177$\arcdeg$ in the $Ks$ band) is consistent
with the orientation of the compact disk
((177$\arcdeg$ $-$ 90$\arcdeg$) = 87$\arcdeg$ $\approx$ 79$\arcdeg$).  
The compact disk around IRS1 must be tilted so that 
the west side is nearer to us 
{if we assume the typical morphological relationship 
between the outflows
and IRNe (e.g., Tamura et al. 1991).}

Our $Ks$ band magnitude of IRS1 is 0.8 mag fainter than 
that in Walther et al. (1993), although $JHKs$ magnitudes
for this source and other sources are roughly consistent with each
other ($<$ 0.5 mag).
%, if we consider the YSO nature of most 
%of these sources.
Note, however, this is not true for IRS5 and several very 
faint magnitudes ($>$ 18 mag).

\subsubsection{IRS2}
IRS2 is not significantly polarized in Walther et al. (1993), 
but in our data it is polarized (7-17\%) at a level of the polarization 
of nearby nebulosity, especially at shorter wavelengths (see Table 1). 
IRS2 positionally coincides with the radio continuum source NGC 2071 1b 
(Snell \& Bally 1986) 
and is considered to be self-luminous. 

There is a local intensity peak and a slight change of
vector pattern to the east of IRS2, but it does not show
a centro-symmetric pattern like IRS1 and IRS8.
Thus the relationship of this local peak and IRS2
in not clear.
{Higher spatial resolution polarimetry is necessary
to reveal these structures.} 
Note however that a local H$_{2}$ emission peak is {associated}
with IRS2 (Aspin et al. 1992). 
We consider that IRS2 is a YSO without extensive circumstellar material.

\subsubsection{IRS3}
IRS3, a red source just north of IRS1, itself is highly polarized ($P$ = 19\% in the $Ks$ band)
and the vector pattern around IRS3 is partly centro-symmteric, 
whose perpendicular lines apparently point toward 
the south-west.

This position is in fact coincident with a radio source NGC 2071 1c (Snell \& Bally 1986).
Because of the limited spatial resolution in both radio and NIR observations,
it is difficult to make absolute astrometry.
However, if we assume IRS2, which is point-like and shows no
centro-symmetric polarization pattern, coincides with the radio source 1b, then
IRS1 coincides with 1a, but IRS3 does not coincide
with 1c. Their positions are plotted in figure 2. 

Therefore, we suggest that IRS3 itself is not self-luminous but is illuminated 
by the radio source 1c, which is situated just south-west of IRS3 and invisible 
even at near-IR wavelengths.
Such suggestions of invisible illuminating sources are
also made in other star forming regions (e.g., GGD27 NIRS2; Tamura et al. 1991) 

The nebula associated with IRS3 is thus an independent, 
compact IRN besides the large IRN associated with IRS1. 
We refer to it as IRN2 (the red IRN in the $PI$ image). 
The levels of polarization degree of IRN2 range from 11\% to 33\% at $Ks$. 
Their northern part, however, has a different pattern;
this part is illuminated probably by IRS1.

Seth et al. (2002) suggested the presence of a nearly edge-on 
circumstellar disk around IRS3 whose axis is in the direction of 35$\arcdeg$. 
The relative position between the radio continuum source and IRS3 
is consistent with the idea that IRS3 is a part of IRN 
illuminated by the invisible radio continuum source. 
If this is the case, the disk around IRS3 is tilted so that 
the south-west side is nearer to us.
Similar to the case of IRS1, the position angle
of the aperture polarization of IRS3 ($\theta$ = 124$\arcdeg$ in the $Ks$ band) 
is consistent with the orientation of the compact disk
((124$\arcdeg$ $-$ 90$\arcdeg$) = 34$\arcdeg$ $\approx$ 35$\arcdeg$).

Eisl{\"o}ffel (2000) suggested that IRS3 is the exciting source
of the H$_{2}$ emission (the flow IIA and IIB), extending
roughly east and west.
However, we consider that their flow IIA is just a part of
the outflow associated with IRS1 and
the flow IIB is with IRS8.
IRN1 and IRN4 (see \S3.2.8) are their counterpart IRNe, respectively.
In fact, their narrow-band H$_{2}$ emission images might suffer from the continuum
emission; our imaging polarimetry of these components
demonstrates that IRS3 (or its illuminating radio source 1c; see above) 
is illuminating none of these.
Aspin et al. (1992) could not see any convincing evidence for activity centered on
IRS3 in their Fabry-Perot H$_{2}$ imaging data.
  
\subsubsection{IRS4}
IRS4 itself is not highly polarized as found by Walther et al. (1993) 
and ours, but our data suggest a {clear} association of a compact reflection nebula 
{with a centrosymmetric pattern} around IRS4, extending to the south-east. 
We refer to it as IRN3. 

Walther et al. (1993) suggested IRS4 to be a background star. 
However, we consider this to be a member of the NGC2071IR cluster 
because of the association of the local nebula IRN3.
IRS4 is also associated with faint H$_{2}$ emission (Aspin et al. 1992).

\subsubsection{IRS5}
IRS5 seems to be a local peak of the IRN1 associated with IRS1. 
Walther et al. (1993) suggested IRS5 as well as IRS5a and IRS8a 
to be shocked H$_{2}$ peaks.
Our $Ks$ and $H$ band magnitudes for IRS5 are more than 0.5 mag
fainter than those in Walther et al. (1993).

\subsubsection{IRS6}
We found that IRS6 is a binary with a separation of 2$''$ and is not highly polarized. 

\subsubsection{IRS7}
IRS7 is a rather isolated red-color source, 
and it is relatively highly polarized ($\sim$10\%). 
It was suggested to be an independent, local outflow source by Walther et al. (1993).

Our polarization vector map shows no clear centro-symmetric pattern
around IRS7.
However, its position angle of polarization
is completely different from that
of nearby point-like sources (see \S 3.4), which is suggestive of
the circumstellar structure such as disk or envelope (e.g., Tamura et al. 2006).
Note also that CO band emission is detected in IRS7, inferring that
this source is a young stellar object (YSO; Walther, Geballe, \& Robson 1991).
Therefore, we suggest that IRS7 is a YSO with a disk/outflow system.

\subsubsection{IRS8}
%IRS8 is not significantly bright in our image, 
%therefore we consider that this is a variable source. 
%More importantly 
IRS8 is a faint red source to the sout-west of IRS1
suggested to be a YSO (Walther et al. 1993).
Associated is a near-IR nebula which shows a clear centro-symmetric pattern, 
thus we believe IRS8 to be a self-luminous illuminating source. 
We {refer to} the bipolar nebula around IRS8 extending both 
to the north-west and to the south-east as IRN4.

\subsubsection{EAST and WEST}
The nebula ``EAST'' in Walther et al. (1993) is a part of 
the large IRN1 illuminated by IRS1. 
The nebula ``WEST'' in Walther et al. (1993) is a part of 
the IRN4 illuminated by IRS8. 
The small counterpart of this IRN is also seen just to 
the south-east of IRS8. 

\subsubsection{Summary}
We have identified four infrared reflection nebulae 
(IRN1-4) and their illuminating sources
in the core region (the NGC2071IR cluster).
IRN1 is the brightest part of the IR nebula
associated with the large scale molecular outflow.
Near-IR source IRS3 is simply a part of the nebula IRN2.
IRN1, IRN2, IRN3, and IRN4 are illuminated by
IRS1 (radio continuum source 1a),
radio continuum source 1c,
IRS 4, and
IRS8, respectively.
The vector pattern is centro-symmetric around IRS1, 4, and 8,
and partly centro-symmetric around 1c.
No distinct local polarization pattern is seen around
IRS5 or IRS6.
IRS7 has an intrinsic polarization, suggestive of
an unresolved disk/outflow system.

All the features discussed above are most clearly seen
in the $Ks$ band but are very similar 
to those seen in our $H$ band vector image. 
Only a part of these IRNe is seen in our $J$ band image
because of extinction in the cloud. 

%\end{enumerate}

\subsection{H$_{2}$ emission polarization and magnetic field structures}

Another interesting feature of the nebula polarimetry is seen in the $Ks$ band, 
in which the knotty H$_{2}$ emission is significantly polarized and 
their vectors are relatively well aligned with each other. 
Note that our $Ks$ band $I$ image is deep enough that most of the H$_{2}$ emission 
features detected by Eisl{\"o}ffel (2000) with narrow band imaging
are recognized in our $Ks$ band image (red knots in figure 1).

Since the shocked H$_{2}$ emission itself is not polarized {\it in situ}, 
its polarizations are likely due to dichroic absorption by aligned grains 
between the H$_{2}$ emission region and the observer, 
thus tracing the magnetic field structure of the NGC 2071 region. 
Such H$_{2}$ emission polarimetry has been conducted for 
only two other regions so far (Orion, Hough et al. 1986; 
Burton et al. 1991; Chrysostomou et al. 1994; 
DR21, Itoh et al. 1999).

Figures 3 (a) and (b) shows the polarization vector map
and the histogram of the position angle of 
the polarization for the brightest H$_{2}$ knots IA, respectively 
(see Eisl{\"o}ffel 2000).
{Polarization vectors are shown only when their signal-to-noise ratio
is better than 4. Therefore, the position angle error of
each vector is better than 8$\arcdeg$.}
The dominant direction is PA $\sim$ 115$\arcdeg$
(FWHM $\sim$ 40$\arcdeg$), 
which is roughly perpendicular to the outflow direction.

Levels of apparent polarization degrees are relatively higher ($>$ 10\%), but
this is probably due to our low signal-to-noise ratio,
especially for the faint knots.
The brightest knots have polarizations of a few percent, consistent with
the level of dichroic polarization in this region discussed below.
%The northern H$_{2}$ knots V are not bright enough
%to obtain useful polarization data.

Figures 3 (c) and (d) shows the similar images
for the H$_{2}$ knots V.
{Polarization vectors are shown only when their signal-to-noise ratio
is better than 2.}
Because of the weaker surface brightness, the number of available
pixels for useful polarizations are limited, but we can still
see an alignment tendency.
Interestingly, the dominant direction is PA $\sim$ 150$\arcdeg$,
different by $\sim$40$\arcdeg$ from the direction in the knots IA,
but with a large dispersion (FWHM $\sim$ 90$\arcdeg$).
Therefore, the magnetic fields might be slightly bent near these knots.
Deeper H$_{2}$ polarimetry is necessary to confirm this result. 

The southern H$_{2}$ knots (IB) are contaminated by reflection or 
ionized nebulosity that is seen as an arc to the south. 
We believe this part to be illuminated by the central star of NGC2071 
(HDE 290861) rather than by IRS1 because only the edge facing to
HDE 290861 is bright.
Unfortunately, the signal to noise ratio in polarization is too small 
and two competing mechanisms make the polarization pattern too complex 
to discuss either magnetic field structures or reflection structures.
However, it is noteworthy that some of the H$_{2}$ knots IB
also show the polarizations (not shown here) consistent with those in the knots IA.
Thus, the magnetic field direction seems to be roughly the same in the
northern and southern outflow regions.
  
\subsection{Aperture polarimetry and magnetic field structures}

No interstellar polarimetry has been made to decide the magnetic field structures
in this $\sim$1 pc region so far.
Therefore, we measured software aperture polarizations of point-like sources
detected in the field-of-view.
We rejected the sources whose photometric errors are greater than 0.1 mag or
whose polarization is greater than 10\%.
{Note that 
the position angle error of each vector is better than 7$\arcdeg$.}
Such a large polarization is unlikely to be dichroic origin
in this region (see \S3.6).
Also rejected are the bright saturated sources.

A total of 53, 79, and 73 sources are measured in the $J$, $H$, and $Ks$ bands, respectively.
The aperture radius was 3 pixels.
Aperture polarization vector maps ($Ks$) of point-like sources superposed 
on the $I$ images are shown in figure 4. 
The distribution of position angles at $JHKs$ is shown in figure 5. 

Although the dispersion is relatively large at shorter wavelengths, 
there is a distinct trend of their position angles;
they are well aligned with each other and considered to
be of dichroic origin (e.g., Tamura et al. 1987), which suggests 
a large scale magnetic field over the observed area (on $\sim$1 pc scale). 
The dominant field direction is PA $\sim$ 120$\arcdeg$ with some deviation 
in the north-east region of the observed field.
Note that the polarization of the central source of IRN6 (see figure 1b)
has a clear intrinsic polarization, which is almost
perpendicular to the polarization of other sources.

There is a slight shift of the peak angle from $Ks$ band to $J$ band 
by $\sim$10$\arcdeg$.
The dispersion is larger at shorter wavelengths; the FWHMs of the histogram
are $\sim$20$\arcdeg$, $\sim$30$\arcdeg$, and $\sim$50$\arcdeg$
in the $Ks$, $H$, and $J$ bands, respectively. 
The origin of this wavelength dependence of position angles
is unknown at present.

This direction of the aperture polarization
is consistent with that derived from the H$_{2}$ polarimetry 
(PA=115$\arcdeg$). 
Both of these are due to dichroic absorption in the NGC 2071 region and 
give the magnetic field direction projected on the sky. 
The only difference is that the aperture polarimetry (of background stars) 
traces the entire cloud along the line-of-sight, while the H$_{2}$ polarimetry 
traces the front-half of the cloud. 
Since these directions are consistent with each other, both 
the aperture polarimetry of field stars and the H$_{2}$ emission polarimetry 
suggest that the magnetic field runs at PA=120$\arcdeg$ over the outflow region, 
which is roughly perpendicular to the large scale outflow direction ($\sim$45$\arcdeg$). 
This direction is somewhat different from the very large scale field 
in the Orion region (Appenzeller 1974) but rather consistent with 
the direction near the OMC-1 region ($\sim$120$\arcdeg$; Kusakabe et al. 2006).

%{\bf Theoretical implication of miss-aligned field (with Matsumoto).}
\subsection{Alignment of outflow/disk systems with magnetic fields}
%According to Matsumoto \& Tomisaka (2004), such a miss-alignment between
%the local magnetic field and the outflow direction ($\sim$75$\arcdeg$)
%suggests that the field strength is relatively weak ($<$ 20 $\mu$G).
Alignment of an outflow with the magnetic field is discussed in
Matsumoto \& Tomisaka (2004).
%\citet{Matsumoto04}. 
They demonstrated a misaligned outflow, which is
ejected in the direction of the magnetic field in the $10-100$~AU
scale, while its direction is inconsistent with the magnetic filed in
the cloud scale.  The alignment is controlled by the magnetic braking
during the protostellar collapse.  The magnetic braking time is
estimated as (see eq.~[29] in 
Matsumoto \& Tomisaka 2004),
%\cite{Matsumoto04}),
\begin{equation}
t_b = \left(\frac{8}{5}\right)^{1/3} \frac{\lambda_J}{v_a},
\end{equation}
where $\lambda_J = c_s(\pi/G\rho)^{1/2}$ and $v_a = B/(4\pi
\rho)^{1/2}$ denote the Jeans length and the Alfv\'en speed,
respectively.  The protostellar collapse is characterized by the free
fall time, defined as,
\begin{equation}
t_\mathrm{ff} = \left(\frac{3\pi}{32 G\rho}\right)^{1/2}.
\end{equation}
The ratio between above two time scales is expressed as a ratio
between the sound speed and the the Alfv\'en speed,
\begin{equation}
\frac{t_b}{t_\mathrm{ff}} = \left(\frac{8}{5}\right)^{1/3}
\left(\frac{32}{3}\right)^{1/2} \frac{c_s}{v_a}
=3.82 \, \frac{c_s}{v_a}.
\label{eq:ratio}
\end{equation}
%\citet{Matthews02} 
Matthews et al. (2002)
estimated the magnetic field strength by examined
their {submillimeter} data using the method of 
Chandrasekhar \& Fermi (1953), and
%\citet{Chandra53}, and
obtained the field strength of $B \approx 56\,\mu$~G, assuming the rms
velocity of $\delta v = 0.5\,\mathrm{km}\,\mathrm{s}^{-1}$ and the
density of $\rho = 3.8\times 10^{-20}\,\mathrm{g}\,\mathrm{cm}^{-3}$
(the number density of $n = 10^{4}\,\mathrm{cm}^{-3}$).  
For the sound
speed $c_s$ in equation~(\ref{eq:ratio}), we should adopt the value of
the rms velocity $\delta v$, which is the so-called effective sound
speed, and it includes the effects of the internal gas motion such as
turbulence.  Applying these values, equation~(\ref{eq:ratio}) yields
$t_b/t_\mathrm{ff} = 2.4$.  This ratio indicates that the magnetic
braking does not have enough time to align the magnetic field of the
protostar scale with respective to the cloud scale magnetic field in
the period of the protostellar collapse. In other words, the magnetic
field is too weak to align the outflow with the magnetic field 
on the large scale.

Although the ratio, $t_b/t_\mathrm{ff}$, estimated above is a somewhat
marginal value, it seems to be under-estimated.  This is because
the method of 
Chandrasekhar \& Fermi (1953)
%\citet{Chandra53} 
has a possibility of over-estimation of
the magnetic field strength as pointed out by 
%\citet{Ostriker01}, \citet{Padoan01}, and \citet{Matsumoto06}. 
Ostriker et al. (2001), Padoan et al. (2001), and Matsumoto et al. (2006).
If the magnetic field strength is weaker than $56\,\mu$~G, the ratio,
$t_b/t_\mathrm{ff}$, has a larger value.

This is in contrast to the OMC-1 region and the NGC7538 region where 
the new born massive YSOs bend the local magnetic field near YSOs 
but the outflows tend to align in the outer ($\sim$0.5 pc) regions 
(Momose et al. 2001).

The weakness of the magnetic field is also exhibited in the position
angles of the aperture polarization of YSOs, since they are indicators
of the local disk/outflow systems (see \S3.2). 
Their position angles represent the position angle of disk, perpendicular
to the outflow direction.
{The aperture polarizations of IRS1, IRS3, IRS7, and IRS8 have 
the position angles
which are different with each other} (see Table 1).  This large dispersion in the
position angle also indicates that the magnetic braking is ineffective
in aligning
of the axes of YSOs in this region.

However, it should also be noted that in clustered star forming regions such as the
NGC 2071 region, 
the direction of the disk/outflow
systems could be affected by the other YSOs, which might
result in a ``random'' distribution of the disk/outflow direction.
A systematic wide-field polarization survey toward various clustered star forming
regions will be required for understanding the relationship between
the YSOs' disk/outflow alignment with the local magnetic field.

\subsection{Near-IR vs. submillimeter polarimetry}

It is interesting to compare the {magnetic} field direction
derived from dichroic polarization at near-IR wavelengths
and that from thermal emission polarization at submillimeter wavelengths
because Goodman et al. (1995) claimed that the former was not always
a good tracer of the magnetic fields {\it within} the dense clouds.
They suggested a possible {inefficiency} of grain alignment
in a dense region.

Figure 6 shows a comparison of the PA histogram in the $H$ band and 
that at the wavelength of 850 $\micron$ (Matthews et al. 2002). 
The 850 $\micron$ emission traces the magnetic field structure 
associated with the dense cloud of the NGC2071IR region 
(covering about 2$'$ = 0.23 pc in diameter), while
the near-IR data cover a $\sim$1 pc scale region.
The general trend of the two directions shows a good agreement with each other,
although their spatial scale is somewhat different. 
Therefore, our results demonstrate that both near-IR dichroic polarization
and submillimeter emission polarization trace 
the magnetic field structures of $\sim$0.1-1 pc scale 
associated with the NGC2071IR region. 

The agreement of the magnetic field structures seen at near-IR 
and far-IR/submillimeter wavelengths is also revealed in the OMC-1 region 
(Houde et al. 2004; Kusakabe et al. 2006), the Ophiuchius dark cloud 
(Tamura et al. 1996), and the NGC 2024 region (Kandori et al. 2006b)

\subsection{Polarization efficiencies}

{Figure 7 shows the degrees of polarization in the $H$ band vs. 
the $H-Ks$ color.} 
There is a weak correlation as is expected from the dichroic polarization (\S3.4). 
{The average slope $P(H)/(H-Ks)$ is 6.0{$\pm$}4.8}, which is consistent 
with that ($\sim$5.6) in the M42 region (Tamura et al. 2006; 
Kusakabe et al. 2006). 
Therefore, the polarization efficiencies have a similar value in 
the two distinct regions in the Orion cloud complex.
Detailed comparison of the polarization efficiency 
with other star forming regions will be 
presented in a future paper.

\section{Conclusion}
We have conducted deep $JHKs$ imaging polarimetry of a 7$\farcm$7 $\times$ 7$\farcm$7 area 
of the NGC 2071 star forming region. Main conclusions are summarized as follows.

\begin{enumerate}

\item Our polarization data have revealed various infrared reflection 
nebulae (IRNe) associated with the central IR young star cluster NGC2071IR and 
identified their illuminating sources. 
There are at least 4 IRNe in NGC2071IR.
IRN1 is the brightest part of the IR nebula
associated with the large scale molecular outflow.
IRN1, IRN2, IRN3, and IRN4 are illuminated by
near-IR source IRS1 (radio continuum source 1a),
radio continuum source 1c,
near-IR source IRS 4, and
near-IR source IRS8, respectively.
We have suggested that IRS3 is simply a part of
IRN2 illuminated by a (near-IR invisible) radio source NGC 2071 1c. 
Four additional IRNe (IRN5-8) are identified around nearby young stars.
The optical reflection nebula, NGC 2071 itself, is not distinct
in the near-IR polarized intensity images.

\item Aperture polarimetry of IRS1 and IRS3 shows
that their polarization position angles are both consistent
with the orientations of the compact ($\sim$10 AU) disks 
traced by water maser emission.

\item Aperture polarimetry of 
the point-like sources within the field in the $JHKs$ bands is
made in this region for the first time.
The magnetic fields 
%(from 7$'$=0.8 pc down to 10$''$=0.02 pc 
%[or, a few arcsec $\sim$ 0.01pc]) 
(from $\sim$0.1 pc to $\sim$1 pc)
derived from the dichroic polarization
are running at a position angle of $\sim$120$\arcdeg$
projected on the sky.

\item Imaging polarimetry in the $Ks$ band shows
the dominant knotty nebulae
of the shocked H$_2$ emission is also polarized.
They are due to dichroic absorption within the NGC 2071 core region.
The derived magnetic field direction projected on the sky
($\sim$115$\arcdeg$) is consistent with that from the polarimetry
of point-like sources.

\item  The magnetic field direction derived from the near-IR polarimetry
is also consistent with that inferred from the previous 850 $\mu$m thermal polarimetry.

\item The magnetic fields in the NGC 2071 region of 0.1-1 pc scale 
are running roughly perpendicular to the direction of the large scale outflow. 
We argue that 
the field strength is too weak to align the outflow in the large scale field direction
via magnetic braking.

\end{enumerate}

%\subsubsection{Subsubsection}
%
%The resent result from ...........

%\newpage
%
%\section{Section 4}
%
%The final ..........
%
%
%%%%%%%%%%%%%%%%%%%%%%%%%%%%%%%%%%%%%%%
%
%We are grateful to Misato Fukagawa for several use-
%ful comments on an earlier version of this paper.
This research was partly supported by MEXT,
Grant-in-Aid for Scientiffc Research on Priority Areas,
``Development of Extra-solar Planetary Science'',
and by grants-in-aid from MEXT (Nos. 16340061, 17204012).
%We thank 
{We thank the referee, David Weintraub, for his helpful comments.}

%\section{Approximation of ...}
%
%\section*{Complete data}
%
%%%
% See the manual for the detail.
%%%

\begin{figure}[t]
    \centering
    \begin{minipage}[c]{1.00\columnwidth}
        \centering\includegraphics[width=\columnwidth]{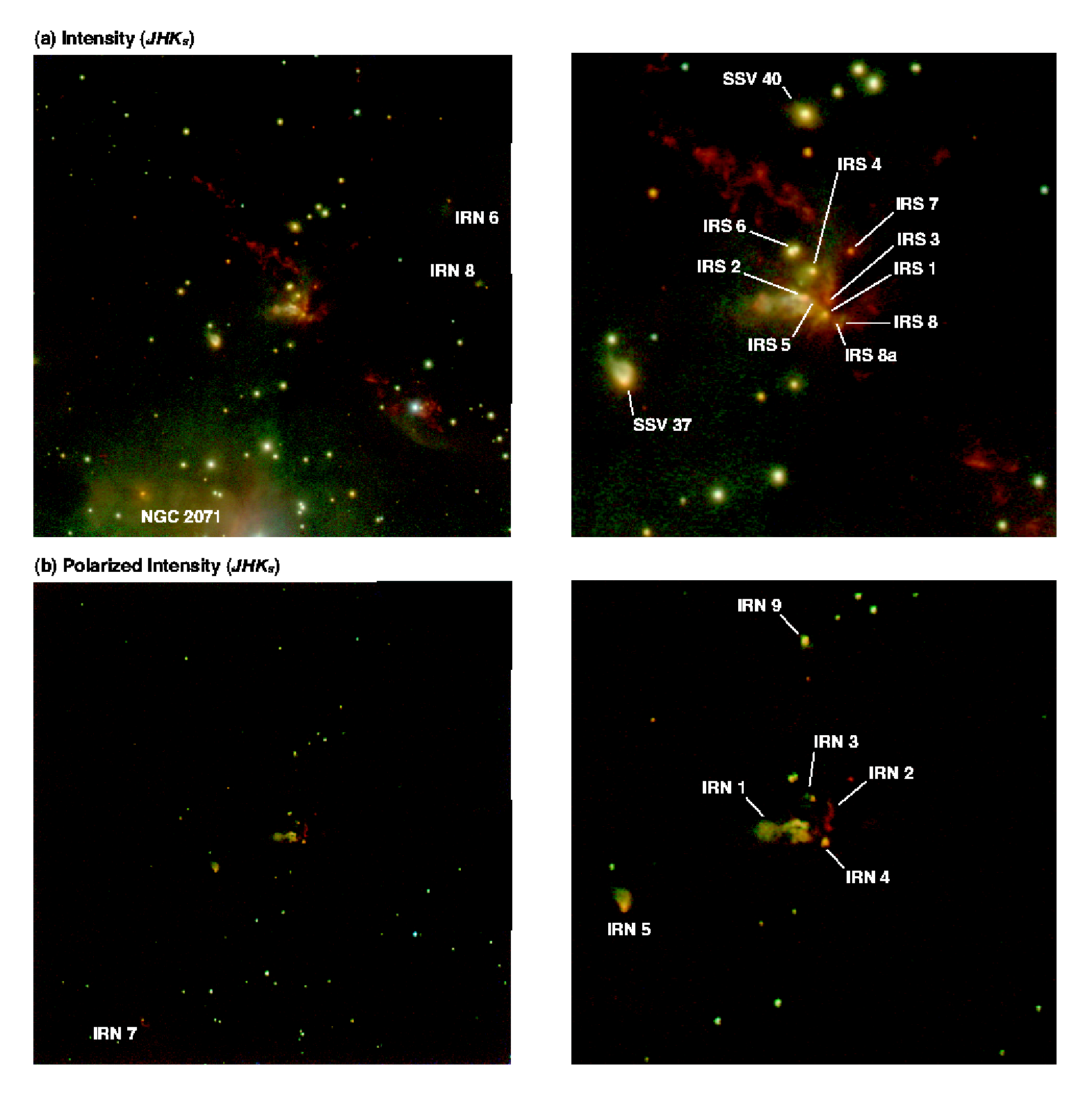}\\
    \end{minipage}
\end{figure}
  
\begin{figure}[t]
%\pagebreak
\clearpage
\centering
    \begin{minipage}[c]{1.00\columnwidth}
        \centering\includegraphics[scale=0.8]{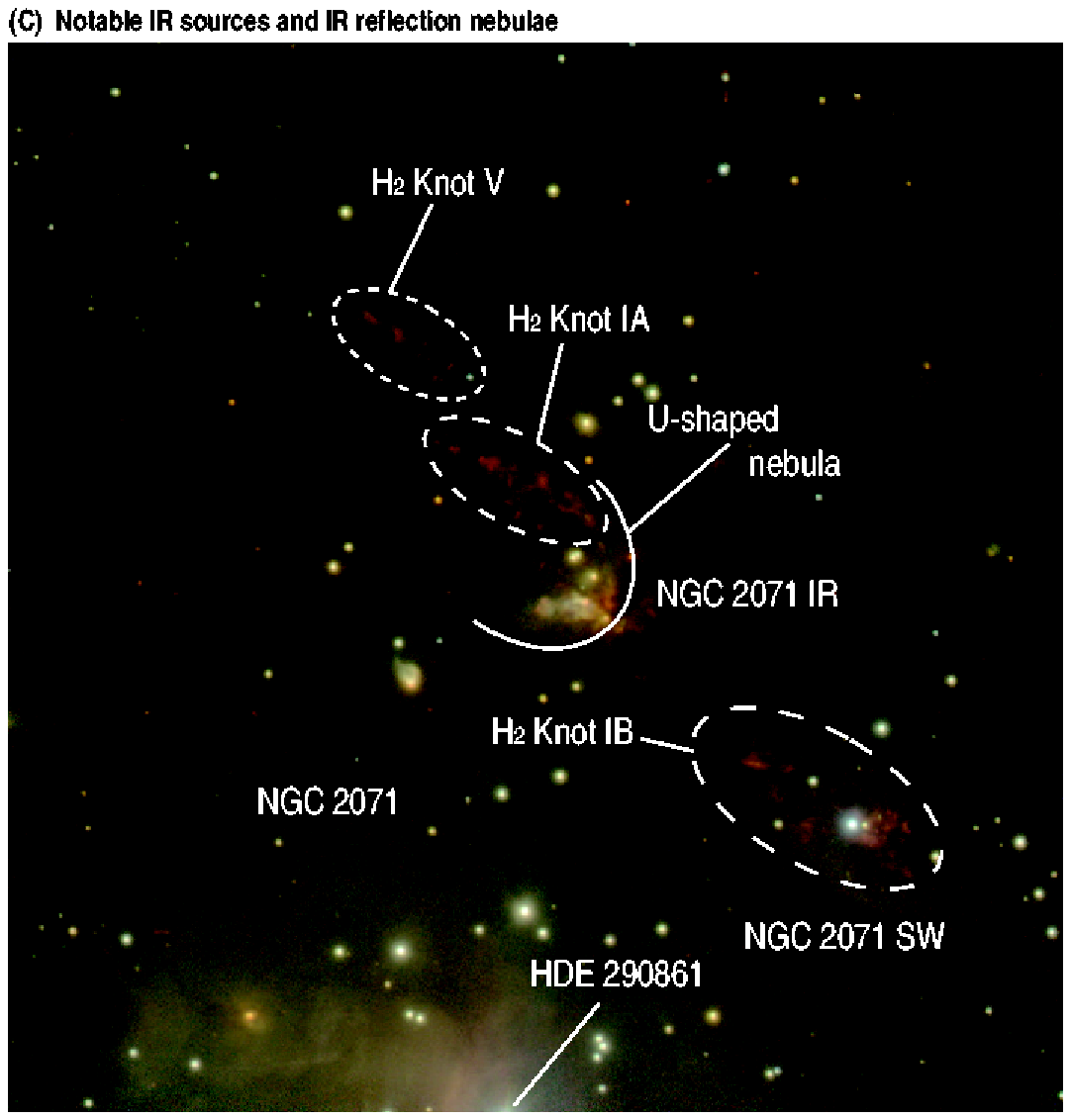}\\
    \end{minipage}
  
%>>>> use \label inside caption to get Fig. number with \ref{}
\caption
{\label{fig:1}$JHKs$ color composite images of intensity 
(a) and polarized intensity (b) of the NGC 2071 region.
$J$ is blue, $H$ is green, and $Ks$ is red.
Field-of-view is 7$\farcm$7 $\times$ 7$\farcm$7 in the left images, while it is 
4$'$ $\times$ 4$'$ in the right images.
(c) Notable IR sources and IR reflection nebulae (IRNe) are indicated.
{Field-of-view is 7$\farcm$7 $\times$ 7$\farcm$7.}}
\end{figure}

\twocolumn

% /Figure/2a-d/ %
\clearpage
\onecolumn
\begin{figure}[t]
    \centering
    \begin{minipage}[c]{1.00\columnwidth}
        \centering\includegraphics[scale=0.45]{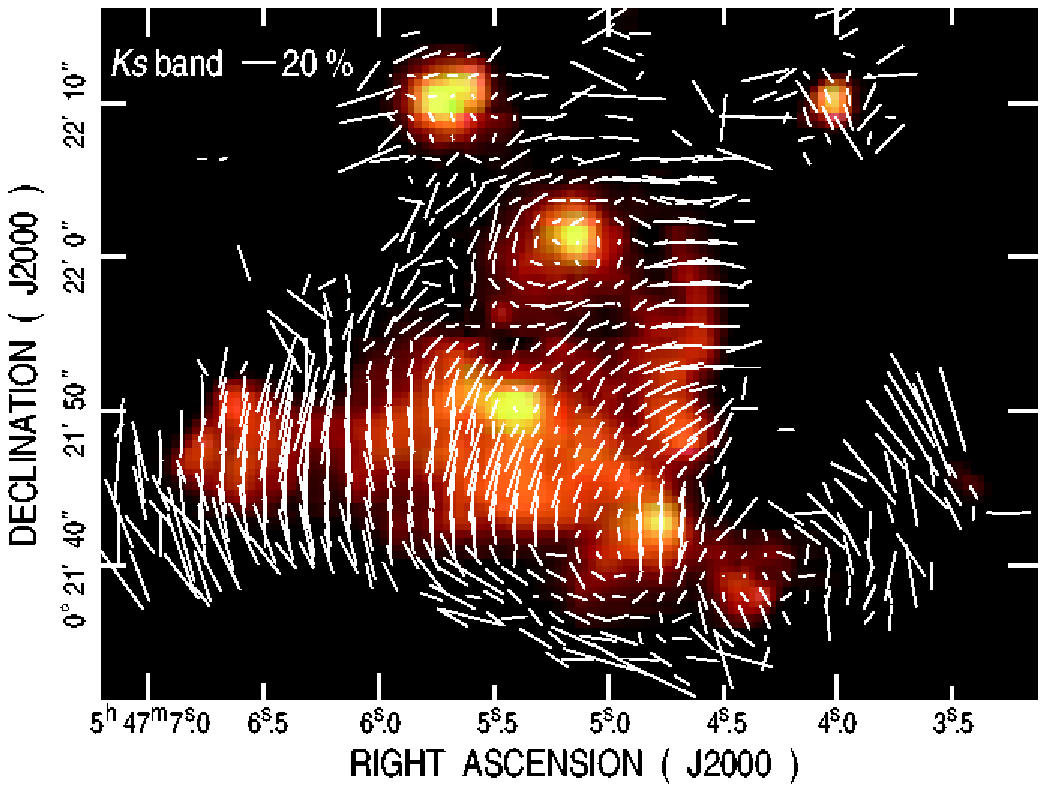}\\
    \end{minipage}
    \centering
    \begin{minipage}[c]{1.00\columnwidth}
        \centering\includegraphics[scale=0.45]{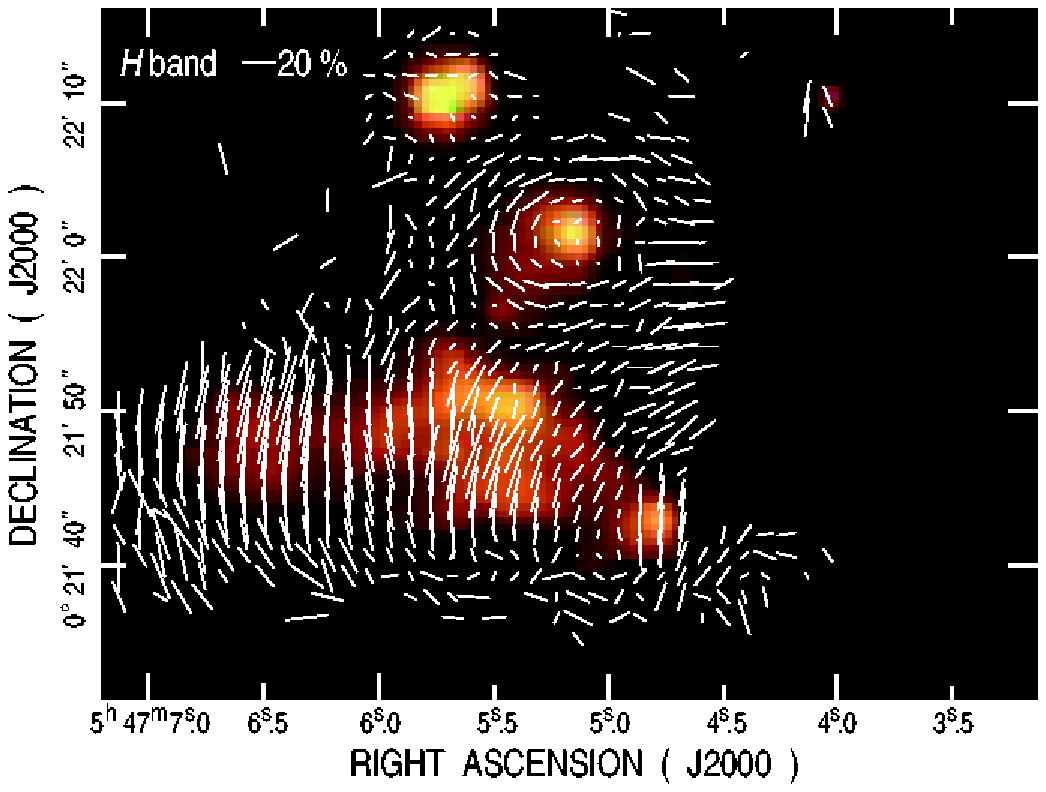}\\
    \end{minipage}
    \centering
    \begin{minipage}[c]{1.00\columnwidth}
        \centering\includegraphics[scale=0.45]{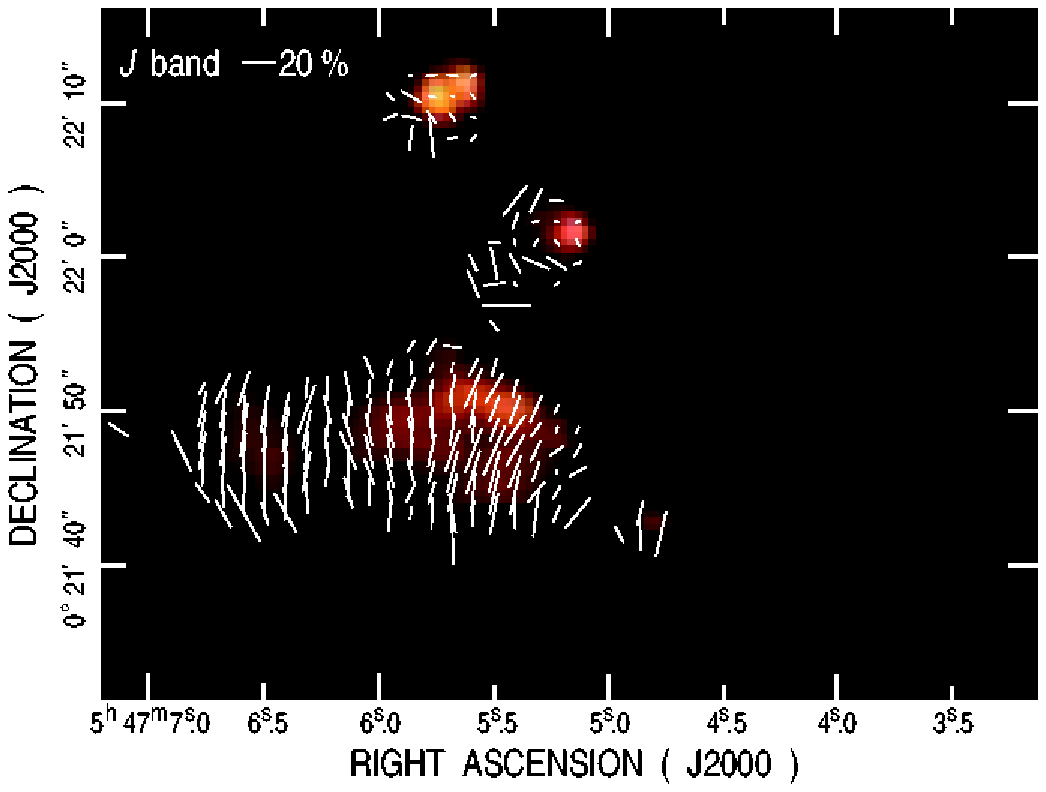}\\
    \end{minipage}
    \centering
    \begin{minipage}[c]{1.00\columnwidth}
%        \centering\includegraphics[scale=0.25]{fig2d.eps}\\
         \centering\includegraphics[scale=0.45]{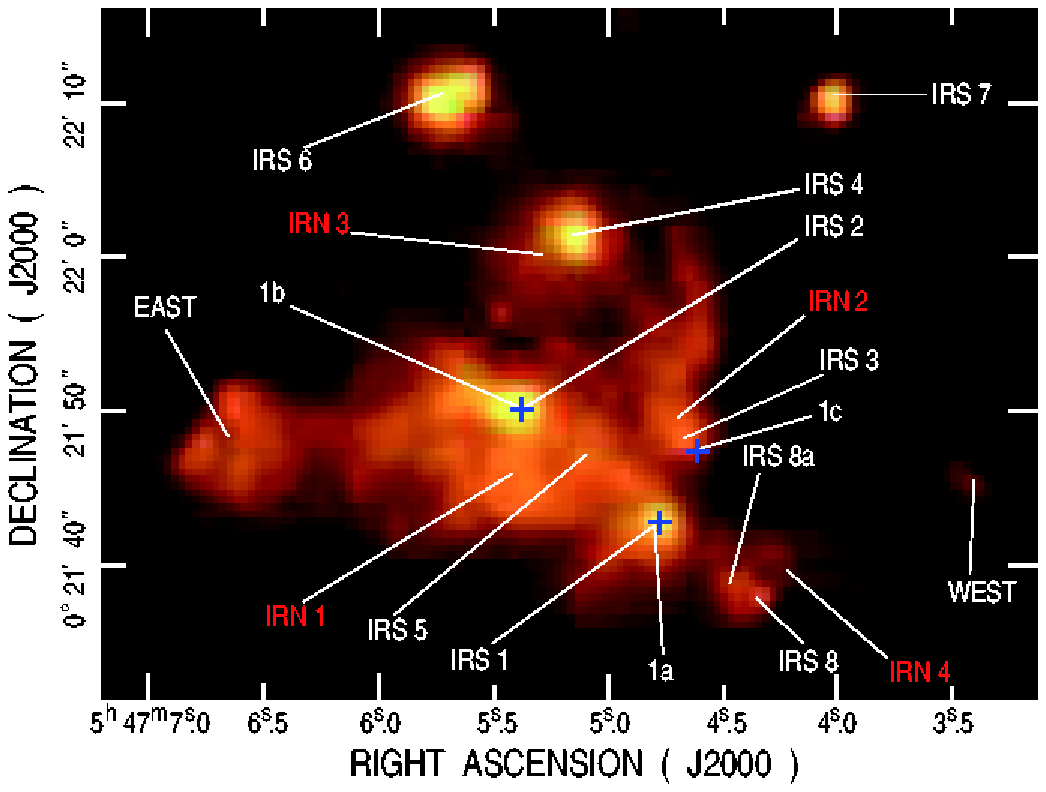}\\
    \end{minipage}
   \caption
%>>>> use \label inside caption to get Fig. number with \ref{}
{\label{fig:2}$KsHJ$ band polarization vector maps and $Ks$ band identification map
of the central N2071IR cluster region superposed on the intensity images.
Notable IR sources and IR reflection nebulae (IRNe) are identified.
Crosses denote the positions of three radio sources.}
\end{figure}
\twocolumn

% /Figure/3/ %
\clearpage
\onecolumn
\begin{figure}[t]
    \centering
    \begin{minipage}[c]{1.00\columnwidth}
        \centering\includegraphics[scale=0.8]{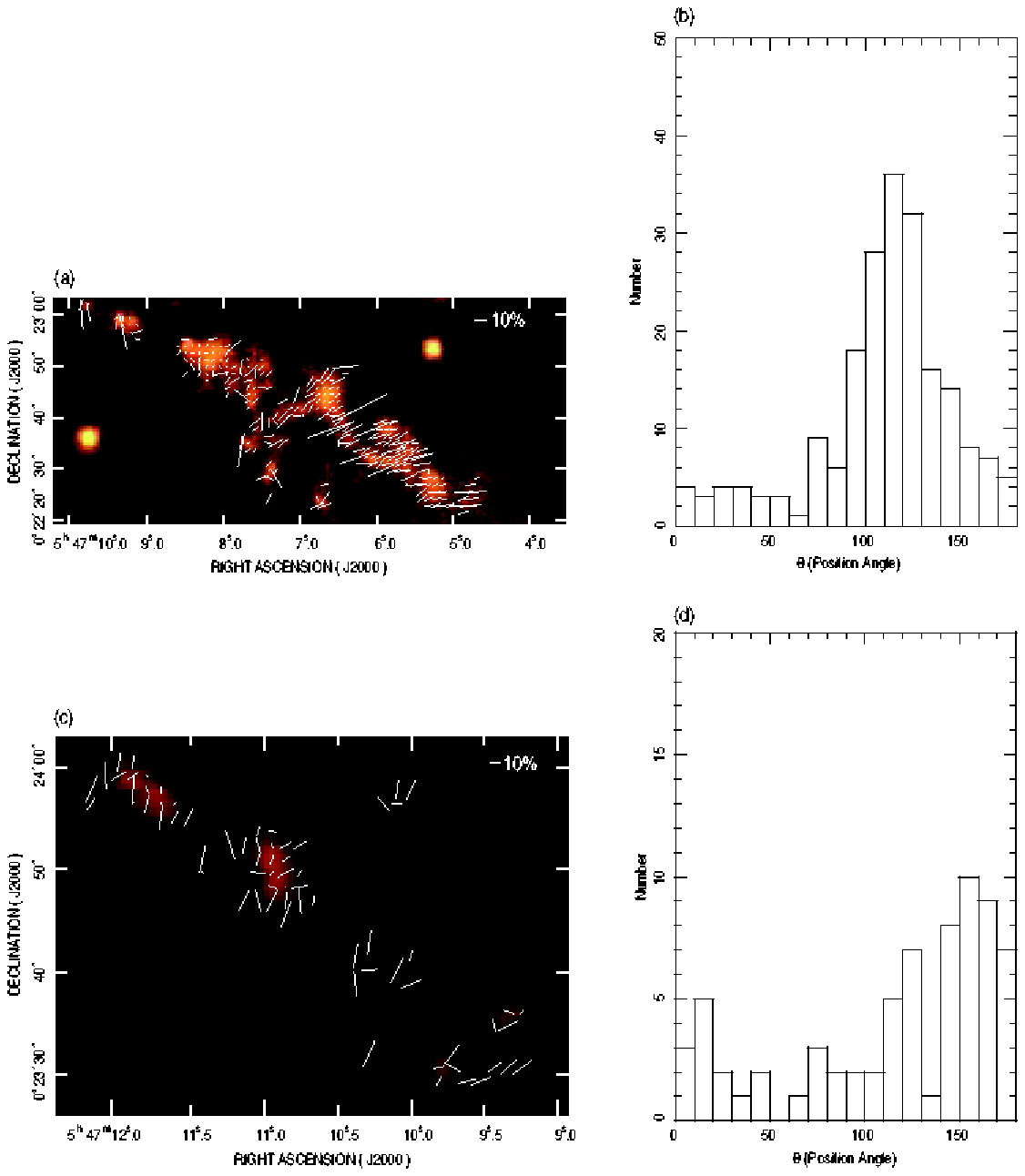}\\
    \end{minipage}

%    \centering
%    \begin{minipage}[c]{1.00\columnwidth}
%        \centering\includegraphics[scale=0.25]{fig3b.eps}\\
%    \end{minipage}
%\clearpage    
%    \centering
%    \begin{minipage}[c]{1.00\columnwidth}
%        \centering\includegraphics[scale=0.45]{fig3c.eps}\\
%    \end{minipage}
%
%    \centering
%    \begin{minipage}[c]{1.00\columnwidth}
%        \centering\includegraphics[scale=0.25]{fig3d.eps}\\
%    \end{minipage}
% 
   \caption
%>>>> use \label inside caption to get Fig. number with \ref{}
{\label{fig:3} (a and b) $Ks$ band polarization vector superposed on the intensity image
of the molecular hydrogen knots IA, and 
%1$''$ corresponds to 10\% of the polarization vector length
%in this figure.
histogram of $Ks$ band polarization position angle
of the molecular hydrogen knots IA.
(c and d) Those for the knots V.}
\end{figure}

\twocolumn

% /Figure/4/ %
\clearpage
\onecolumn
\begin{figure}[t]
    \centering
    \begin{minipage}[c]{1.00\columnwidth}
        \centering\includegraphics[width=\columnwidth]{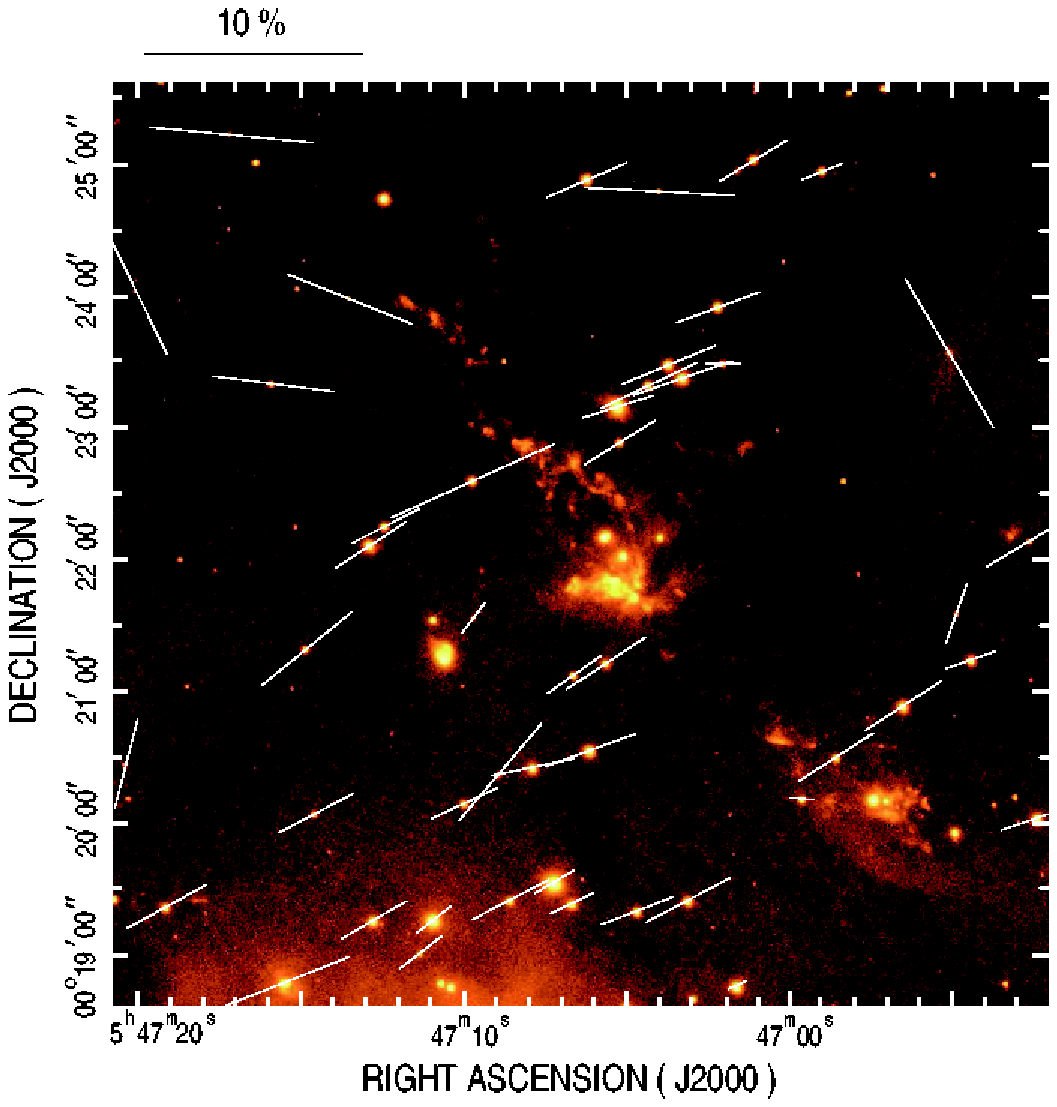}\\
    \end{minipage}
   \caption
%>>>> use \label inside caption to get Fig. number with \ref{}
{\label{fig:4}$Ks$ band polarization map
of point-like sources superposed on the intensity image.
Note that the polarization of the central source of IRN6
seen in the north-west, has a clear intrinsic polarization, which is almost
perpendicular to the polarization of other point-like sources.}
\end{figure}

\twocolumn

% /Figure/5/ %
\clearpage
\onecolumn
\begin{figure}[t]
    \centering
    \begin{minipage}[c]{1.00\columnwidth}
        \centering\includegraphics[scale=0.95]{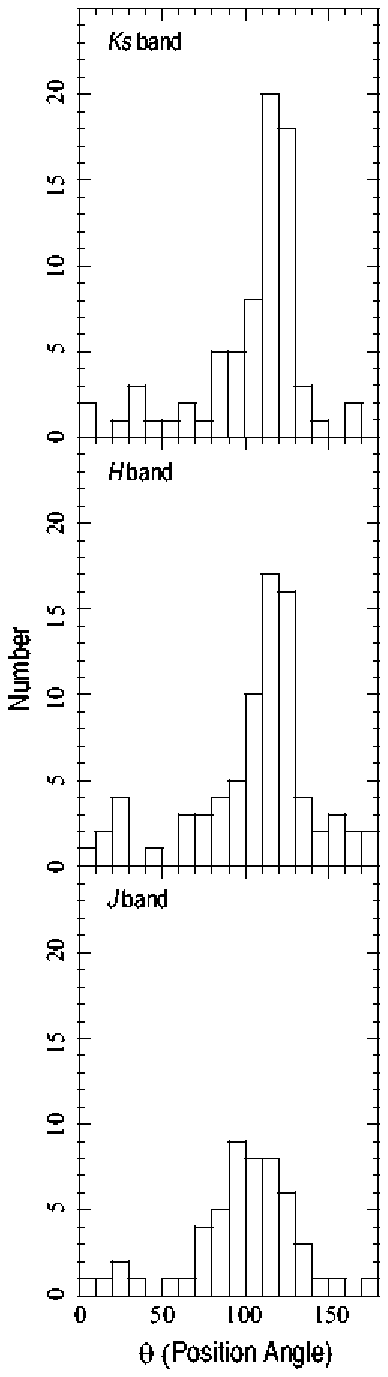}\\
    \end{minipage}
   \caption
%>>>> use \label inside caption to get Fig. number with \ref{}
{\label{fig:5}$KsHJ$ band histograms of position angle
of point-like sources.}
\end{figure}

\twocolumn

\twocolumn

% /Figure/6/ %
\clearpage
\onecolumn
\begin{figure}[t]
    \centering
    \begin{minipage}[c]{1.00\columnwidth}
        \centering\includegraphics[scale=0.7]{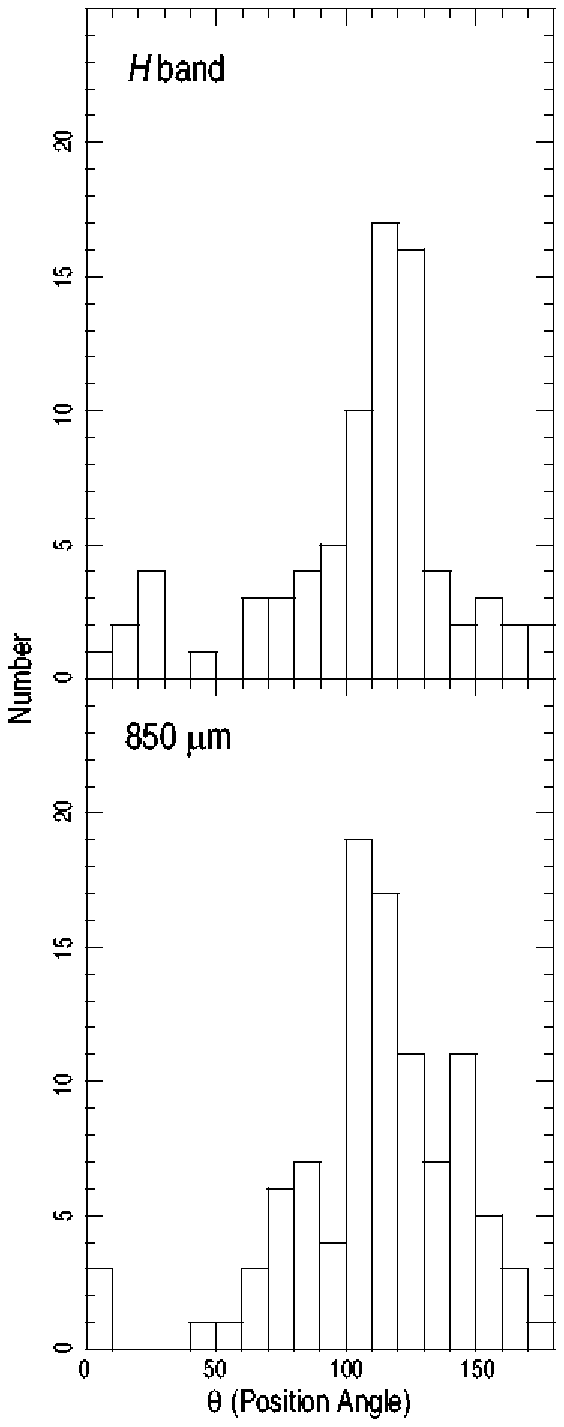}\\
    \end{minipage}
   \caption
%>>>> use \label inside caption to get Fig. number with \ref{}
{\label{fig:6}Comparison of the distribution of position angles
at 1.63 $\mu$m and 850 $\mu$m. The 850 $\mu$m data are rotated by 90$^{\circ}$.}
\end{figure}

\twocolumn

% /Figure/7/ %
\clearpage
\onecolumn
\begin{figure}[t]
    \centering
    \begin{minipage}[c]{1.00\columnwidth}
        \centering\includegraphics[scale=0.7]{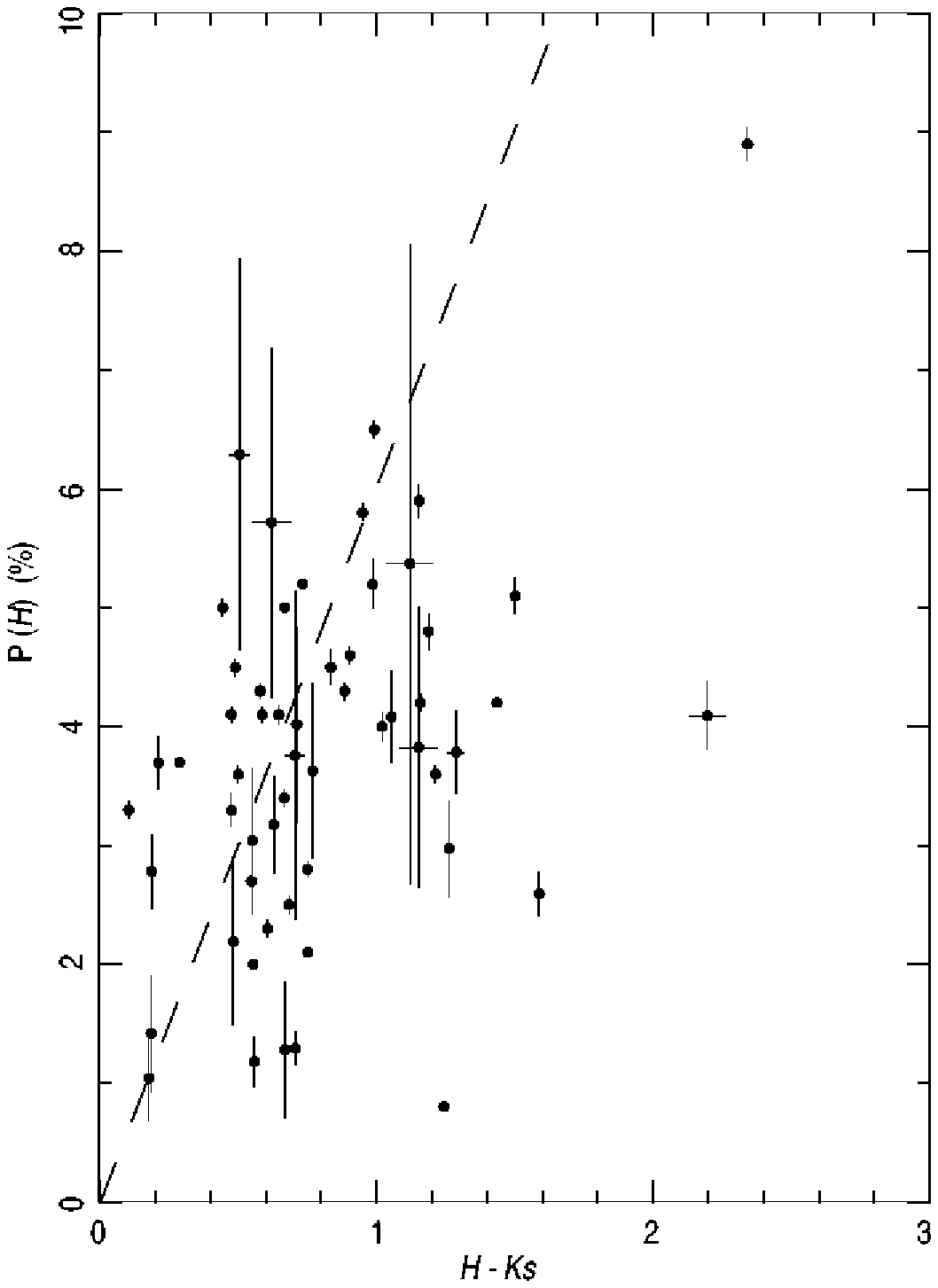}\\
    \end{minipage}
   \caption
%>>>> use \label inside caption to get Fig. number with \ref{}
{\label{fig:7}Correlation between degrees of polarization
in the $H$ band and $H-Ks$ color. {The broken line shows
a linear fit with a slope of 6.0.}}
\end{figure}

\twocolumn

\end{document}